# Two successive magneto-structural transformations and their relation to enhanced magnetocaloric effect for $Ni_{55.8}Mn_{18.1}Ga_{26.1}$ Heusler alloy

Zhe Li[1,*], Kun Xu[1], Yuanlei Zhang[1], Chang Tao[1], Dong Zheng[2], Chao Jing[2]

[1]College of Physics and Electronic Engineering, Key Laboratory for Advanced Functional and Low Dimensional Materials of Yunnan Higher Education Institute, Qujing Normal University, Qujing 655011, P. R. China

[2]Department of Physics, Shanghai University, Shanghai 200444, P. R. China

Corresponding Author: [*]Zheli@shu.edu.cn, +868748968627

**Abstract** In the present work, two successive magneto-structural transformations (MSTs) consisting of martensitic and intermartensitic transitions have been observed in polycrystalline $Ni_{55.8}Mn_{18.1}Ga_{26.1}$ Heusler alloy. Benefiting from the additional latent heat contributed from intermediate phase, this alloy exhibits a large transition entropy change $\Delta S_{tr}$ with the value of ~27 J/kg K. Moreover, the magnetocaloric effect (MCE) has been also evaluated in terms of Maxwell relation. For a magnetic field change of 30 kOe, it was found that the calculated value of refrigeration capacity in $Ni_{55.8}Mn_{18.1}Ga_{26.1}$ attains to ~72 J/kg around room temperature, which significantly surpasses those obtained for many Ni-Mn based Heusler alloys in the same condition. Such an enhanced MCE can be ascribed to the fact that the isothermal entropy change $\Delta S_T$ is spread over a relatively wide temperature interval owing to existence of two successive MSTs for studied sample.

During the past decades, Ni-Mn based ferromagnetic shape memory alloys (FSMAs) have been attracted much attention because of their multiplicity of functional properties like magnetic shape memory effect (MSME),[1-2] magnetocaloric effect (MCE)[3-5] and magnetoresistance (MR),[6-7] etc. Among these FSMAs, stoichiometric Ni$_2$MnGa is the most representative, which undergoes two separate transformations containing a magnetic transition at $T_C \approx 376\,\text{K}$ and a first-order martensitic transition (MT) at $T_M \approx 202\,\text{K}$.[8] At the temperature located between $T_C$ and $T_M$, Ni$_2$MnGa is ferromagnetic (FM) and has a cubic $L2_1$-type austenitic structure. Upon cooling from $T_M$, it would transform to a tetragonal martensitic structure maintaining its FM ordering with high magnetocrystalline anisotropy. Besides the MT, a first-order intermartensitic transition (IMT) between the modulated and the unmodulated martensite, caused by changes in composition, temperature and external stress, has been also extensively investigated in Ni-Mn-Ga alloys.[9-14] In comparison to MT, the IMT usually occurs at a much lower temperature, resulting in two entirely separated transformations. Currently, the IMT associated with the change of magnetization has been continuously reported in a new type of FSMAs, such as Ni-Mn-In-Sb,[15] Ni-Co-Mn-Sn[16,17] and Ni-Cu-Mn-Sn.[18] For these alloys, the IMT and MT were mostly found to be closing upon each other and hence form two successive magneto-structural transformations.[16-18] Owing to the sequent magneto-structural couplings, such a kind of the multiple MSTs, compared with the intermediate phase in Ni-Mn-Ga alloys, can show more abnormal physical properties.[16,17]

As is well known, both $T_C$ and $T_M$ for Ni-Mn-Ga alloys are particularly sensitive to their composition. Previous experimental studies conformed that partial substitution of Ni with Mn can decrease $T_C$ and increase $T_M$ in $Ni_{50+x}Mn_{25-x}Ga_{25}$.[19-21] In some particular compositions, the magnetic transition would happen to coincide to MT. Such a coincidence also gives rise to the MST from paramagnetic (PM) austenitic phase to FM martensitic phase near room temperature. This individual behavior is similar to that observed in some giant magnetocaloric materials [22-24] and $Ni_2In$-type hexagonal compounds.[25-27] It makes these particular alloys prospectively apply to magnetic refrigeration.[28-33] Very recently, a giant MCE related to magneto-multistructural transformation was reported in annealed $Ni_{52}Mn_{26}Ga_{22}$ ribbon above room temperature, which is due to existence of an intermediate phase involving different modulated martensitic structures.[34] In the present work, very interestingly, two successive MSTs have been observed in $Ni_{55.8}Mn_{18.1}Ga_{26.1}$ Heusler alloy. Associated with such an exotic behavior, an enhanced MCE around room temperature has been obtained for a magnetic field change of 30 kOe.

## Results

**Thermal magnetization and thermal strain.** The temperature dependence of magnetization for $Ni_{55.8}Mn_{18.1}Ga_{26.1}$, during cooling and heating, is shown in Figure 1. For the sample on cooling, an abrupt change of magnetization appears in the vicinity of $T_M$, which corresponds to a direct MT. With further lowering temperature around $T_{IM}$, the magnetization shows a nonlinear uptrend and gradually attains to a steady stage, which is attributed to a direct IMT. Between the cooling and heating process, there exists two obviously thermal hysteresis for both transformations, which are estimated as $\Delta T_M$=10 K and $\Delta T_{IM}$=8 K, respectively. Meanwhile, a similar jump can be also explored in the thermal strain curve near the same temperatures (see right upper panel of Figure 1). All these findings clearly imply that each transitions are first order and they comprise two successive MSTs within the temperature range of 305 K~335 K. Additionally, the characteristic temperatures of these transformations determined from $dM/dT(T)$ curves (see left lower panel of Figure 1) are equal to $T_M$ =318 K, $T_{IM}$ =313 K, $T_A$ =328 K, and $T_{IA}$ =321 K, where the $T_M$ , $T_{IM}$ , $T_A$ , and $T_{IA}$ denote the direct MT/IMT and the reverse MT/IMT equilibrium temperature, respectively.

**Heat flow, specific heat and crystal structure of different martensitic phases.** To further investigate the transforming behaviors, the heat flow data were collected for $Ni_{55.8}Mn_{18.1}Ga_{26.1}$ by continuous heating and cooling, as shown in Figure 2. During cooling, the direct MT and IMT are accompanied with the well-defined peaks on the heat flow due to the latent heat of these transitions. During heating, the endothermic curve almost displays an identical feature, but it can be distinguished that the reverse IMT slightly weakens. From the inset of Figure 2, it is conspicuous that the $C_p$(T) curve measured with cooling mode also displays two exothermic peaks, which are similar to those observed in heat flow data. Crucially, the corresponding temperatures of these peaks are in good agreement with the characteristic temperatures determined from the magnetic measurements (see Figure 1), which provide an evidence on the occurrence of the two successive MSTs in present sample. According to previous experimental works,[10, 12-14] the sequence of IMT observed in the cooling process is usually from a five-layered (5M) to a seven-layered (7M) modulated martensitic structure or from 7M to non-modulated martensitic structure ($L1_0$). To clarify the sequence of IMT for $Ni_{55.8}Mn_{18.1}Ga_{26.1}$ alloy, Figure 3 shows the x-ray diffraction pattern at room temperature. The reflections indicate that the sample crystallizes into the non-modulated tetragonal ($L1_0$) martensitic structure. The refinement result reveals that the sample possesses the lattice parameters with $a = b = 0.7659$ nm, $c = 0.6644$ nm and $c/a = 0.8675$. After heating the sample near the finish temperature of reverse MT (326 K), the obvious splitting of (222) peak suggests that the present sample is predominately in the 7M modulated martensitic structure with small residual traces of non-modulated martensitic structure (see the inset of Figure 3). Analogous structure (mixed martensitic phase) has also been reported in $Ni_{2.14}Mn_{0.84}Ga_{1.02}$ caused by IMT.[12] Hence, such an experimental result directly proves that the IMT for studied sample should follow with the sequence of 7M to $L1_0$ on cooling and the state returns with thermal hysteresis to the 7M phase on heating.

**Transition entropy change, thermal cycles and Isothermal magnetization.** For $Ni_{55.8}Mn_{18.1}Ga_{26.1}$ alloy, by using heat flow data, the calculated absolute latent heat, |$\Delta L$|, contributed from two sequent transformations is ~8.6 kJ/kg upon cooling and ~8 kJ/kg upon heating, respectively. The small discrepancy in amounts can be ascribed to the presence of degenerative reverse IMT (see Figure 2). However, it is worth noting that both values of $\Delta L$ are larger than that reported in $Ni_{55}Mn_{20}Ga_{25}$ single crystal,[28] on account of the additional latent heat from the IMT. Furthermore, the entropy as a function of temperature computed by exothermic curve from heat flow data, as plotted in Figure 4(a). One can notice that the transition entropy change ($\Delta S_{tr}$) obtained during cooling amounts to ~27 J/kg K, which is comparable to those of giant magnetocaloric materials.[22-24] As shown in Figure 4(b), more importantly, it highlights that the multiple transformations are found to be nicely reproducible after a number of thermal cycles, exhibiting an intrinsic nature. These results predict that the studied sample may be a promising candidate for magnetic refrigeration.

In addition to a large $\Delta S_{tr}$, another vital feature for magnetocaloric materials is the field-induced transformation. With this aim we carried out measurements of isothermal magnetization at selected temperatures for $Ni_{55.8}Mn_{18.1}Ga_{26.1}$, as shown in Figure 5. Both of the representative hysteresis curves are presented at 305 K and 325 K between field-up and field-down courses, indicating a strong FM state of martensite and a magnetically inhomogeneous state coexisting in ferromagnetic and paramagnetic regions of austenite with the temperature close to transition point. At the mediate temperature of 321 K, interestingly, it can be clearly explored that the ascending branch shows a nonmonotonic tendency in the magnetizing process at an inflection point of about 7.5 kOe, which was determined as a peak value on the $dM/dH$ curve. (see the inset of Figure 5). In the subsequent demagnetizing stage, the descending branch manifests a typical FM feature accompanied with a distinct magnetic hysteresis, indicating that there only exists one-way field-induced MT. The same phenomenon was also observed in Ni-Mn-Ga with similar composition.[28,35] In general, the changes of phase volume fraction caused by isothermal magnetic field mainly rely on $dT_M/dH$, which can be assessed by Clausius-Clapeyron (C-C) equation, $|dT_M/dH| = |\Delta M/\Delta S_{tr}|$. Utilizing the values of $\Delta M$ (~27.3 emu/g, see Figure 5) and $\Delta S_{tr}$ for $Ni_{55.8}Mn_{18.1}Ga_{26.1}$, the calculated value of $dT_M/dH$ is only about 0.1 K/kOe. Therefore, the irreversible field-induced MT can be attributed to an insufficient magnetic field, which cannot overcome more thermal energy generated by phase boundary friction.[36] Despite the $dT_M/dH$ of $Ni_{55.8}Mn_{18.1}Ga_{26.1}$ is significantly lower than that in a great deal of metamagnetic materials,[2,5,22-27] we still deem that an enhanced MCE associated with two successive MSTs should be expected.

**Isothermal entropy change and refrigerant capacity.** To confirm this point as mentioned above, we now turn our attention to discussion of MCE for $Ni_{55.8}Mn_{18.1}Ga_{26.1}$, as shown in Figure 6. The inset of this figure depicts the $M(T)$ curves measured at various magnetic fields upon cooling. It can be detected that the effect of magnetic field on the two successive MSTs is negligible except that the transformation regions shift towards higher temperature. The increasing rate is in well correspondence with the result calculated by C-C equation. Based on these $M(T)$ curves, the isothermal entropy change ($\Delta S_T$) during direct MST was calculated under different magnetic fields through the Maxwell relation $(\partial M/\partial T)_H = (\partial S/\partial H)_T$ (see Figure 6). Due to the existence of intermediate phases for $Ni_{55.8}Mn_{18.1}Ga_{26.1}$, it can be found that the MCE occurs in the two steps around room temperature, corresponding to the MT and IMT, respectively, bringing about two sequent $\Delta S_T$ peaks with the same sign. These behaviors are analogous to that reported for high-pressure annealing Ni-Co-Mn-Sn alloys.[16,17] With increasing applied magnetic field, two sequent peaks move to higher temperatures and develop in amplitude simultaneously, which reflect the intrinsic nature of MCE during the multiple transformations. Moreover, the refrigerant capacity (RC), which is the other pivotal parameter for judging materials' magnetocaloric capability, has been estimated by integrating the area under $\Delta S_T$ (*T*) curve from T$_1$ to T$_2$ (shade area in Figure 6). In the case of present sample, although the value of $\Delta S_T$ is only ~11 J / kg K, the RC still achieves ~72 J/kg for the change of magnetic field from 0 to 30 kOe. Such a RC is strikingly higher than that in many Ni-Mn based Heusler alloys[3,4,16,28-34] and comparable to some Ni$_2$In hexagonal-type magnetocaloric compounds in the uniform condition.[37-39] This is attributed to be resulted from two successive MSTs, which can broaden the range of work temperature and the resulted MCE will thus be enhanced dramatically.

**Discussion**

Up to date, a transformation from PM austenitic phase to FM martensitic phase has been developed in some materials like Ni-Mn based Heusler alloys [19-21,28-35] and Ni$_2$In hexagonal-type compounds,[25-27,38-39] etc. Such a MST can be understood by the fact that the $T_C$ of martensitic phase may be higher than that the temperature of the MT occurred. In the case of our studied sample, as distinct from aforementioned materials, Figure 1 and 2 indicate that it experiences two successive MSTs consisting of a MT (from PM to weak FM state) and an IMT (from weak FM to strong FM state). According to previous studies, [9,14,18] we consider that the origin of the multiple transformations could be related to chemical stress (change in composition) or twining stress. In contrast to similar transforming features reported in polycrystalline Ni-Mn-In-Sb[15] and Ni-Cu-Mn-Sn[18] as well as Ni-Mn-Ga ribbons[34], the other appealing aspect in present sample is that, even if an applied magnetic field is up to 30 kOe, the multiple transformations are still persisted, showing a excellently thermodynamic stability (see the inset of Figure 6). Such an outstanding performance of studied sample is a consequence of the fact that the application of magnetic field tends to stabilize its martensitic phase that possesses a higher magnetization in comparison to its austenitic phase. Since the both transformations are joined together strongly, the magnetic field can not only induce entropy change contributed from MT but also produce an additional part contribution of latent heat involved in the IMT (see Figure 5). Consequently, the present sample reveals an enhanced MCE associated with the two successive MSTs (see Figure 6) and appears to be potential candidate for magnetic refrigeration.

In summary, the transformation properties in $Ni_{55.8}Mn_{18.1}Ga_{26.1}$ Heusler alloy have been studied systematically. Our results sufficiently demonstrated that there are two successive magneto-structural transformations in the sequence of austenite→7M martensite→$L1_0$ martensite in the process of cooling. Accompanied by this exotic property, an enhanced MCE has been obtained around room temperature when the magnetic field changes from 0 to 30 kOe. These experimental findings can also help us to develop more efficiently magnetic refrigerants with such a kind of Heusler alloy system.

**Methods**

Polycrystalline $Ni_{55.3}Mn_{19.7}Ga_{25}$ alloy with nominal composition was fabricated from high purity Ni, Mn, Ga elements, by using conventional arc-melting in an argon atmosphere. The weight loss after melting was found to be less than 1%. For homogenization, the obtained ingot was annealed in an evacuated quartz capsule for 72 hours at 1073 K, and slowly cooled to room temperature. Its real composition was determined by energy-dispersive spectrometer (EDS, ProX, Phenom) analysis to corresponding to $Ni_{55.8}Mn_{18.1}Ga_{26.1}$. The crystalline structure at different temperature was identified by x-ray diffraction using Rigaku Ultima-IV x-ray diffractometer. Both of magnetization and specific heat were characterized by physical property measurement system (VersaLab, Quantum Design). Heat flow data were collected by differential scanning calorimeter (DSC, Q2000, TA) on modulated mode with a cooling/heating rate of 3 K/min, and the thermal cycles were also performed by this equipment with a cooling/heating rate of 10 K/min. Examination of thermal strain using a rectangular specimen with a dimension of $2\times10\times10$ mm$^3$, was acquired in standard strain-gauge technique.

## Acknowledgments

This work was supported by the National Natural Science Foundation of China (Grant Nos.11364035, 11404186 and 51371111), the Key Basic Research Program of Science and Technology Commission of Shanghai Municipality (Grant No. 13JC1402400), and Project for Innovative Research Team of Qujing Normal

University (Grant No.TD201301), and Project for Applied Basic Research Programs of Yunnan Province (Grant No. 2013FZ110 and 2012FD051).

**Author contributions**

The idea was proposed by Z.L and the experiments were carried out by Y.L.Z, C.T and D.Z. The experimental results were analyzed and interpreted by Z.L, K.X, Y.L.Z, and C.J. The manuscript was written and corrected by Z.L, K.X and C.J. All authors reviewed the manuscript.

**Additional information**

**Competing financial interests:** The authors declare no competing financial interests.

**Figure captions**

**Figure 1**  Temperature dependence of magnetization $M(T)$ for $Ni_{55.8}Mn_{18.1}Ga_{26.1}$ alloy on cooling and heating at magnetic field of 500 Oe. Left lower panel shows the $dM/dT(T)$ curves corresponding to these transformations. Right upper panel shows the strain as a function of temperature $\Delta l/l(T)$ in the absence of magnetic field upon cooling.

**Figure 2**  Heat flow data measured by continuous heating and cooling for $Ni_{55.8}Mn_{18.1}Ga_{26.1}$ alloy. The inset shows temperature dependence of specific heat $C_p(T)$ during cooling for this sample.

**Figure 3**  X-ray diffraction pattern for $Ni_{55.8}Mn_{18.1}Ga_{26.1}$ alloy at room temperature. Inset: X-ray diffraction pattern at 326 K for this sample directly heated from room temperature.

**Figure 4**  (a) The entropy as a function of temperature during the multiple transformations at zero fields in $Ni_{55.8}Mn_{18.1}Ga_{26.1}$ alloy. (b) Heat flow data of thermal cycling with 21 times around the multiple transformations for this sample.

**Figure 5**  Isothermal magnetization curves for $Ni_{55.8}Mn_{18.1}Ga_{26.1}$ measured at selected temperatures. Prior to measurements, the sample was first warmed in zero fields to pure austenite and then cooled to measuring temperature. The inset shows a peak value on the $dM/dH(H)$ curve measured at 321 K. Arrows indicate the directions of magnetic field change.

**Figure 6**  Temperature dependence of isothermal entropy change $\Delta S_T$ for $Ni_{55.8}Mn_{18.1}Ga_{26.1}$ alloy at different magnetic fields. The inset shows temperature dependence of magnetization under various magnetic fields for this sample.

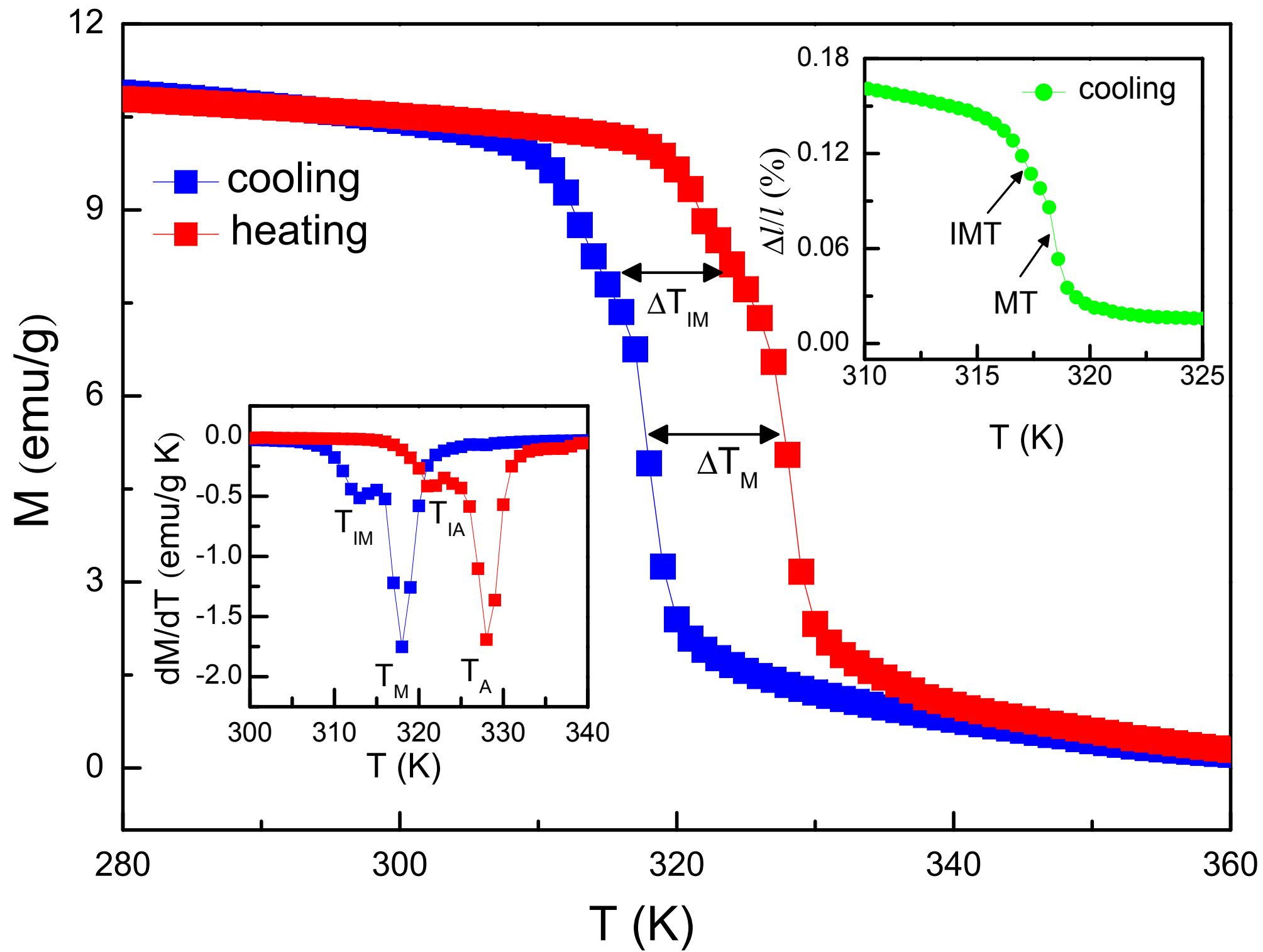

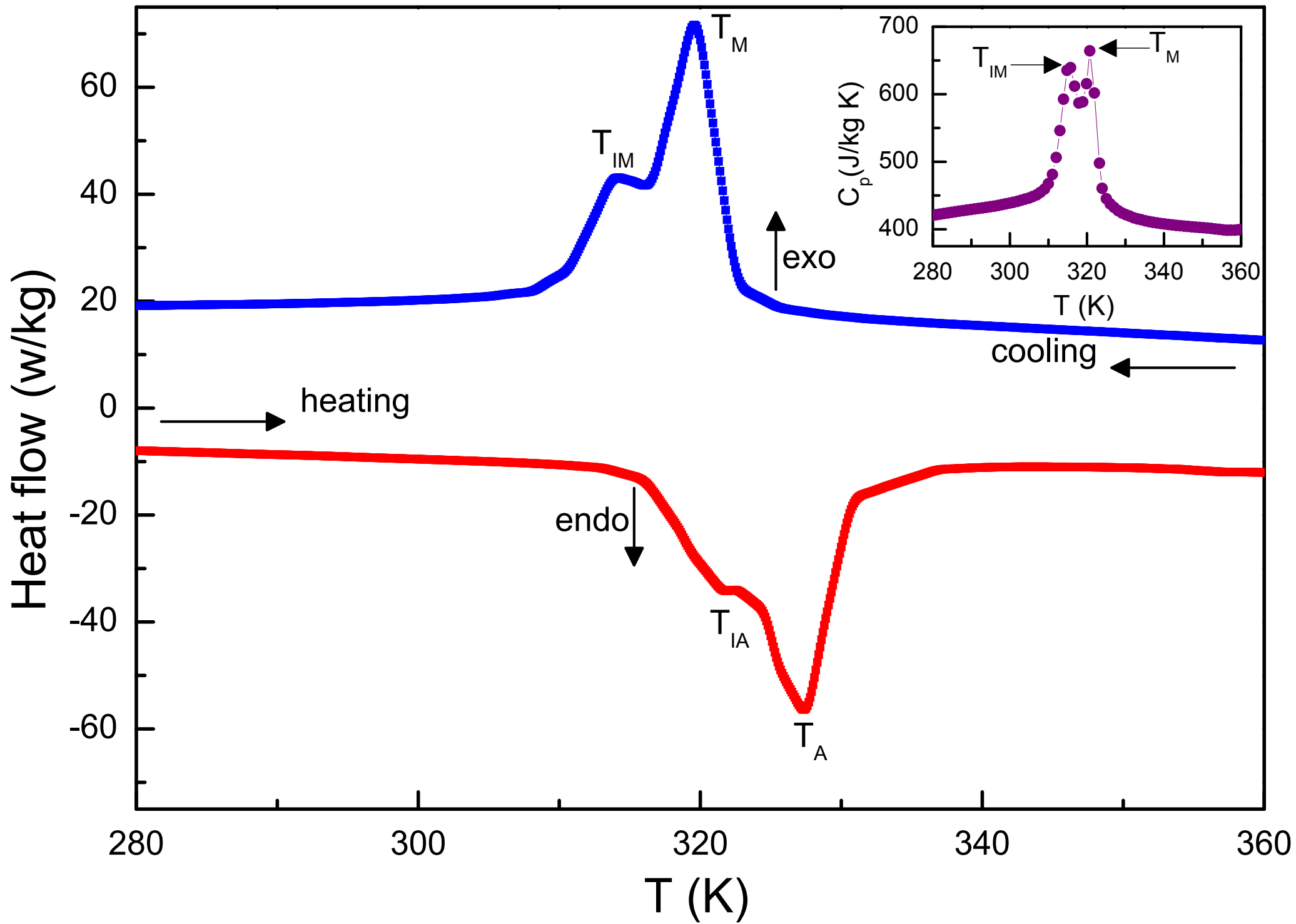

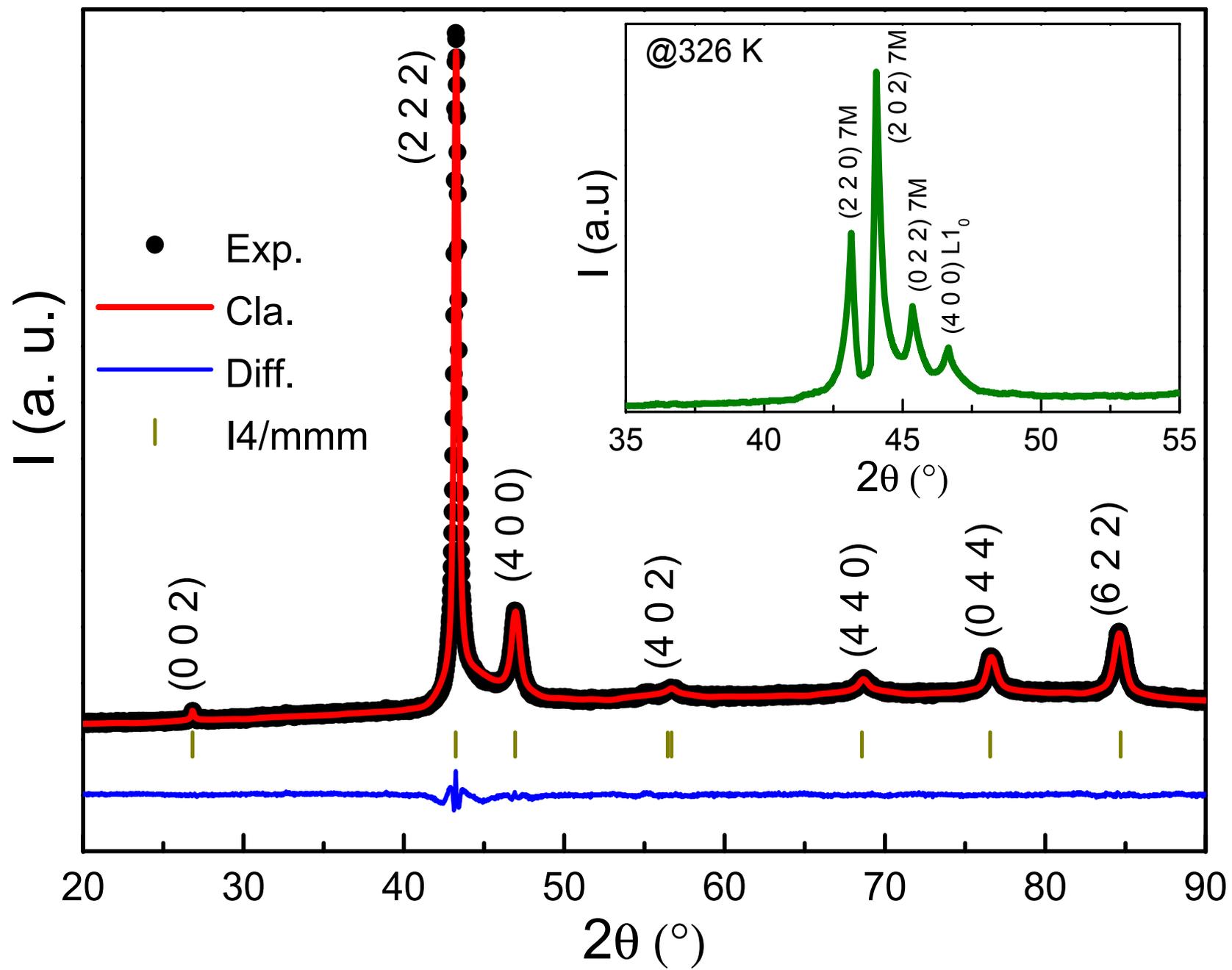

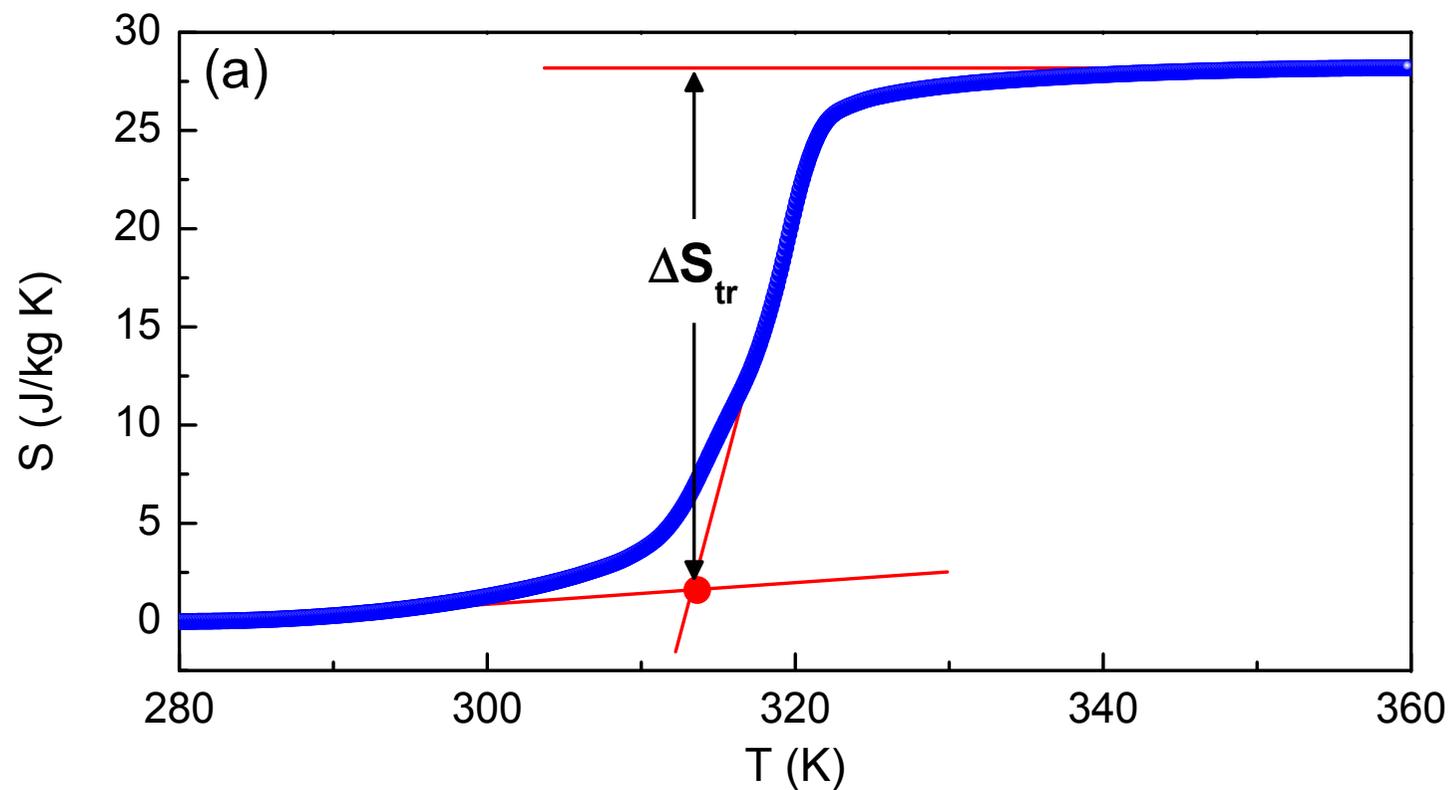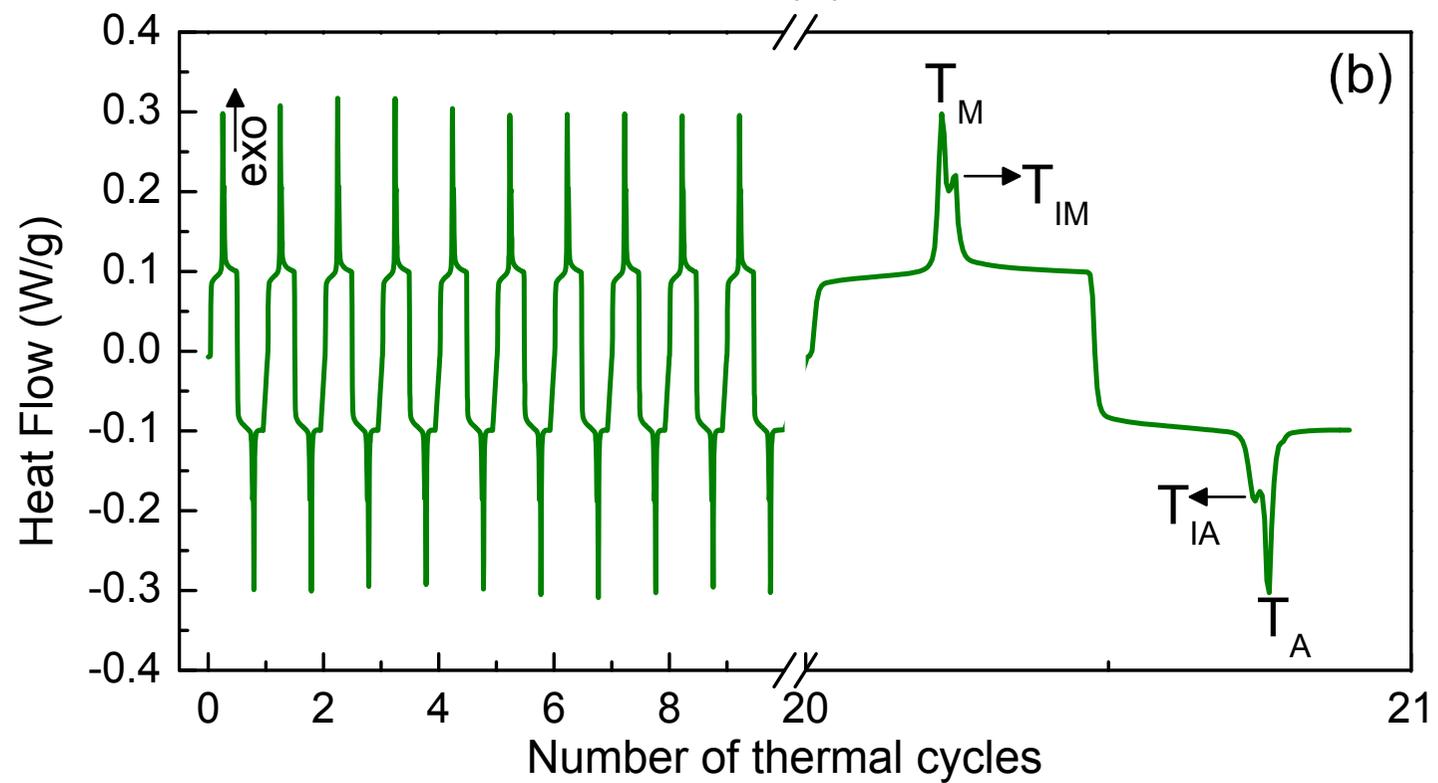

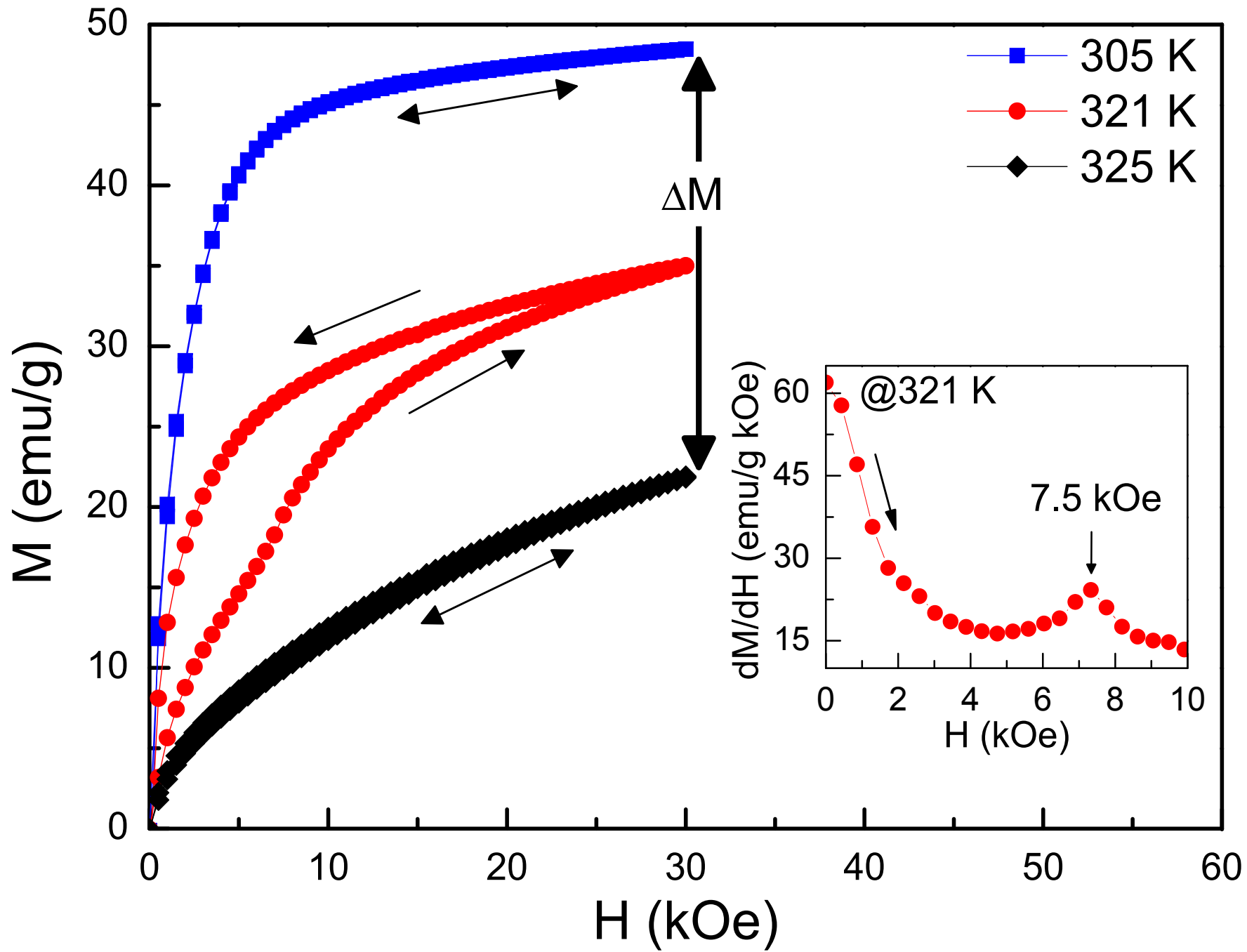

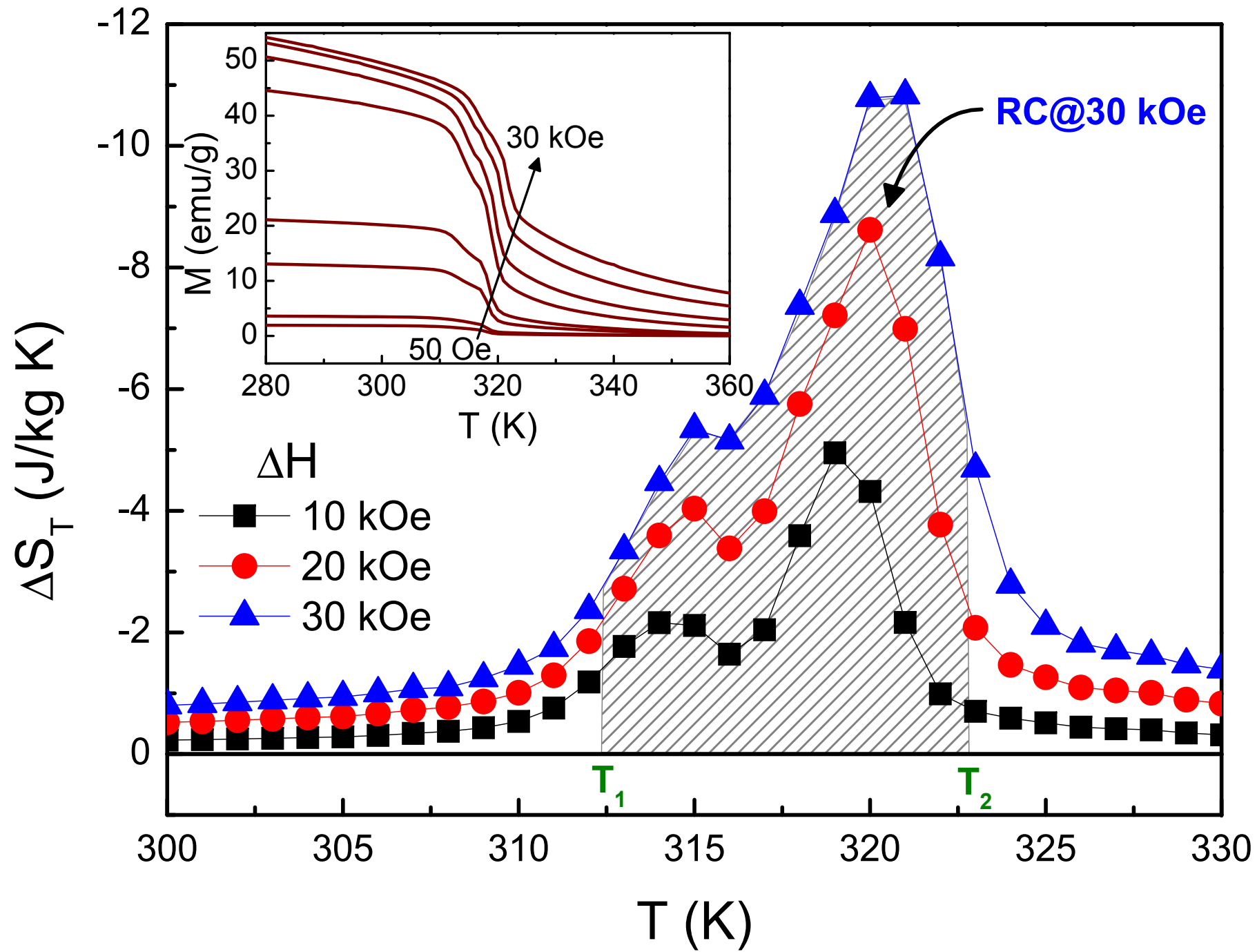